# SOLUTION OF HYDRAULIC FRACTURE PROBLEM ACCOUNTING FOR LAG

Alexander M. Linkov

*Institute for Problems of Mechanical Engineering, 61, Bol'shoy pr. V. O., Saint Petersburg, 199178, Russia*
*Rzeszow University of Technology, ul. Powstancow Warszawy 8, Rzeszow, 35-959, Poland*
*ve-mail: linkoval@prz.edu.pl*

***Abstract.*** The paper presents a method for solving hydraulic fracture problems accounting for the lag. The method consists in matching the outer (basic) solution neglecting the lag, with the inner (auxiliary) solution of the derived 1D integral equation with conditions, accounting for the lag and asymptotic behavior of the opening and the net-pressure. The method refers to practically important cases, when the influence of the local perturbation, caused by the lag, becomes insignificant at a distance, where the leading plane-state asymptotics near the fracture front is still applicable. The universal asymptotics are used for finding the matching constants of the basic (outer) solution and for formulation of matching condition for the solution of inner (auxiliary) problem. The method is illustrated by the solution of the Spence and Sharp plane-strain problem for a fracture propagating symmetrically from the inlet, where a Newtonian fluid is pumped at a constant rate. It is stated that the method developed for deep fractures may also serve in the intermediate range between deep and shallow fractures.

## 1. Introduction

Fluid driven fractures are of significant interest for studying natural geological processes (e.g. Pollard & Johnson 1973; Spence & Turcotte 1985; Lister 1990; Rubin 1995; Roper & Lister 2005), and for various geo-engineering problems (e.g. Penman 1977; Abou-Sayed et al. 1994; Economides & Nolte 2000; Jeffrey & Mills 2000; Murdoch & Slack 2002; Bolson & Cochetti 2003). In view of practical applications to hydraulic fracturing in oil, gas and heat production, there are numerous publications on analytical, approximate and numerical solutions of the problem starting from the pioneering works by Khristianovich and Zheltov (1955), Perkins and Kern (1961), Geertsma and deKlerk (1969), Nordgren (1972), Spence and Sharp (1985). Reviews are given in many papers (e.g. Garagash & Detournay 2000; Adachi & Detournay 2002; Savitski & Detournay 2002; Adachi et al. 2007; Gordeli & Detournay 2011; Garagash, Detournay & Adachi 2011; Linkov 2012; Mishuris et al. 2012; Mitchel et al. 2007; Gordeli & Peirce 2013 and the proceedings of the Conference "Effective and Sustainable Hydraulic Fracturing, HF-2013" (Brisbane, 20-22 May, 2013)).

From the theoretical analysis (e. g. Garagash & Detournay 2000; Garagash 2006), it follows that to have physically consistent picture of the fluid pressure near the fracture contour, there should be a lag between the fluid front and the contour. The published analytical solutions for the lag has referred to asymptotic regimes of either steady propagation of the fracture (Lister 1990; Garagash & Detournay 2000), or the early stage of the propagation (Garagash 2006). The analysis also shows that commonly the lag is several orders of magnitude smaller than a characteristic length of a fracture. The obviously local influence of the lag has justified neglecting it in the most of theoretical studies and computational codes. It would be unreasonable to overcomplicate, say, a code simulating propagation of a hydraulic fracture with a characteristic side of several meters by including a fine mesh capable to catch effects in a near contour strip with the width of some millimeters. Meanwhile, for shallow fractures, the lag is not thus small (Bunger 2005; Lecampion & Detournay 2007, Gordeli & Detournay 2011; Gordeli & Peirce 2013) and it is accounted for by using uniform meshes on the whole fracture.

The intermediate range between deep fractures, for which the lag is negligibly small, and shallow fractures, for which it is significant, has not been addressed to the date. It looks of value, from both the scientific and practical points of view, to have a means to cover this range. There are also other engineering questions regarding the influence of the lag to be addressed. At the conference HF-2013 the possibility to change the fracture toughness of wetted rock by proper control of the lag was discussed. The local effects become also of practical significance when evaluating the influence of proppant bridging, rock inhomogenuities, intersection with natural flaws, evaporation and pore fluid flow near a fracture contour. Thus, besides the scientific significance of studying the zone of local perturbation, comprehensively explained in the paper by Garagash & Detournay (2000), there are engineering problems, which require accounting for the local perturbations caused by the lag when it is moderately or even arbitrary small. Having a general solution for the perturbed zone, we may extend it to the intermediate range of depths.

The regular way to obtain the needed general solution consists in combining an *outer* (basic) solution, neglecting a small lag, with an asymptotic *inner* (auxiliary) solution for steady propagation with the speed defined by the outer solution. The concluding remark of the paper by Garagash & Detournay (2000) mention this option. Still, to the best of our knowledge, it has not been employed so far.

The objective of the present paper is to develop a method which provides the needed matching of the outer (basic) solution and the inner solution of a properly formulated auxiliary problem. We suggest such a method and formulate the condition under which it is applicable. The *basic* solution is found by conventional methods neglecting the lag. Comparing it with the universal asymptotic solution provides two matching parameters: the matching distance $r_m$ and the net pressure at the matching point. The *auxiliary* problem corresponds to the propagation of a semi-infinite fracture with the speed equal to that found from the basic solution at a considered point and time. It is formulated for an arbitrary power-law fluid and power-law leak-off. The comparison of its solution with the universal asymptotic solution provides the minimal distance $r_a$, at which the local perturbation, caused by the lag, may be neglected. It appears that the matching is possible when $r_a$ is less than $r_m$. Then combing the basic and inner solutions provides the complete solution accounting for the lag. The method is illustrated by considering the Spence and Sharp (1985) plane-strain problem for a fracture propagating symmetrically from the inlet, where a Newtonian fluid is pumped at a constant rate. It is stated that the method developed for deep fractures may also serve in the intermediate range between deep and shallow fractures.

## 2. Problem formulation

Consider a fracture with the surface $S$ and the contour $L$, propagating in the 3D space. The fracture is driven by a fluid with the viscosity law of power-type:

$$\tau = M\dot{\gamma}^n, \tag{2.1}$$

where $\tau$ is the shear stress, $\dot{\gamma}$ is the doubled shear strain rate, $M$ is the consistency index, $n$ is the behavior index (for thinning fluids, commonly used in hydraulic fracturing, $0 < n < 1$; for a Newtonian fluid $n = 1$; for a perfectly plastic fluid $n = 0$).

As known (e.g. Spence & Sharp 1985; Garagash & Detournay 2000; Adachi & Detournay 2002; Savitski & Detournay 2002; Adachi et al. 2007; Peirce & Detournay 2008), the system of partial differential equations for the fluid consists of the mass conservation (continuity) equation, which for incompressible fluid reads:

$$\frac{\partial w}{\partial t} + div(w\boldsymbol{v}) + q_l = 0, \tag{2.2}$$

and the Poiseuille type equation for fluid flow in a narrow channel, which for the power viscosity law (2.1) is:

$$\boldsymbol{v} = -\left(\frac{1}{\mu'}w^{n+1}\right)^{\frac{1}{n}}|grad p|^{\frac{1}{n}-1}grad p. \tag{2.3}$$

In equations (2.2) and (2.3), $w$ is the fracture opening, $\boldsymbol{v}$ is the vector of the in-plane particle velocity, $p$ is the pressure, $\mu' = 2\left[\frac{2(2n+1)}{n}\right]^n M$, $q_l$ is the prescribed intensity of distributed sources or sinks of fluid (usually $q_l$ is positive and accounts for leak-off of a fluid into rock formation). For certainty, we shall assume that leak-off is described by the power-type dependence

$$q_l = 2C_l(t - t_*)^{-\beta_l}, \qquad\qquad 0 \le \beta_l < 1, \tag{2.4}$$

where $C_l$ and $\beta_l$ are known constants, $t_*$ is the time instant when the fluid front reaches a considered point. For the Carter's leak-off, $\beta_l = 0.5$; for impermeable rock, $C_l = 0$. Divergence in (2.2), gradient in (2.3) and the vector $\boldsymbol{v}$ in these equations are defined in the plane tangent to the fracture surface at a considered point.

The derivation of the continuity equation (2.1) from the Reynolds transport theorem (e.g. Crowe et al. 2009) employs the fact that for a continuous media, the speed $v_*$ of the front propagation equals the component $v_n$ of the particle velocity normal to the front. This expresses the fundamental speed equation (SE) [Kemp 1990, Linkov 2011, 2012]:

$$v_n = v_* = \frac{dx_*}{dt} \tag{2.5}$$

at each point $x_*$ of a moving fluid front.

In hydraulic fracture problems, the opening $w$ depends on the fluid pressure $p$ in a fracture, fracture geometry, rock stiffness and stresses $\sigma_{ij0}$ in host rock. In the commonly considered case of elastic rock, the dependence is of the form:

$$Aw = p(\boldsymbol{x}, t) - \sigma_0(\boldsymbol{x}), \qquad\qquad \boldsymbol{x} \in S, \tag{2.6}$$

where $A$ is a linear integral operator, $\sigma_0(\boldsymbol{x})$ is the normal component of traction induced by the field $\sigma_{ij0}$ at a given point $\boldsymbol{x}$ of the fracture surface $S$. The integral equation (2.6) is solved under the condition of zero opening at points of the fracture contour $L$, which in general does not coincide with the fluid front $L_f$:

$$w(\boldsymbol{x_L}) = 0, \qquad\qquad \boldsymbol{x_L} \in L. \qquad (2.7)$$

To let the fracture propagate, a fracture condition is imposed at the points of the crack contour $L$. Usually, the condition of linear fracture mechanics is used (see, e.g. Rice 1968):

$$K_I = K_{IC}, \qquad\qquad (2.8)$$

where $K_I$ is the normal stress intensity factor (SIF) at a given point of the contour, $K_{IC}$ is its critical value defined by the strength of rock.

At points of the fluid surface $S_f$, equations (2.2), (2.3) and (2.6) define the opening $w$ and the pressure $p$. Meanwhile, at the lag zone between the contours $L$ and $L_f$, the boundary conditions (BC) are to be specified. They may account for various factors such as evaporation, or the pore pressure in host rock. For certainty, we assume the condition of zero traction on the lag surface:

$$p(\boldsymbol{x}, t) = 0, \qquad\qquad \boldsymbol{x} \in S \cap S_f. \qquad (2.9)$$

By continuity, the BC (2.9) on the lag surface defines also the BC for the fluid equations at points of the fluid front:

$$p(\boldsymbol{x_*}) = 0, \qquad\qquad \boldsymbol{x_*} \in L_f. \qquad (2.10)$$

The problem also requires initial conditions (IC) since the PDF (2.2) contains the first temporal derivative of the opening, while the SE (2.5) involves the temporal derivative of the position of the fluid front. Therefore, we need to prescribe the initial opening $w_0(\boldsymbol{x})$ and the initial position $\boldsymbol{x_{*0}}$ of the fluid front at the initial time $t_0$:

$$w(\boldsymbol{x}, t_0) = w_0(\boldsymbol{x}), \qquad\qquad \boldsymbol{x} \in S_{f0}, \qquad (2.11)$$

$$\boldsymbol{x_*}(t_0) = \boldsymbol{x_{*0}}, \qquad\qquad \boldsymbol{x_{*0}} \in L_{f0} \qquad (2.12)$$

with the initial fluid surface $S_{f0}$ having the contour $L_{f0}$.

The mathematical problem consists of solving the lubrication equations (2.2), (2.3) together with the elasticity equation (2.6) under the IC (2.11), (2.12) and the BC (2.10) for the fluid, the BC (2.7), (2.9) for the lag tip and lag surface and the condition (2.8), defining the size of the lag. The SE (2.5) with $v_*$ defined by (2.3) at the fluid front, serves to trace the changes of the front position in time. We need to find the opening and pressure distribution in both the fluid and lag regions, the position of the fluid front and the lag as functions of time for $t > t_0$.

In many cases, the in-situ traction $\sigma_0$ on the fracture surface is constant or its change is negligible on the scale of the lag-zone. In these cases, it is convenient to use the net-pressure $p_{net} = p - \sigma_0$ instead of the physical pressure $p$. Then the elasticity equation (2.6) and the BC (2.9), (2.10) become, respectively,

$$Aw = p_{net}(\boldsymbol{x}, t), \qquad\qquad \boldsymbol{x} \in S, \qquad (2.13)$$

$$p_{net}(\boldsymbol{x}, t) = -\sigma_0, \qquad\qquad \boldsymbol{x} \in S \cap S_f, \qquad (2.14)$$

$$p_{net}(\boldsymbol{x}_*) = -\sigma_0, \qquad\qquad \boldsymbol{x}_* \in L_f. \qquad (2.15)$$

From now on, we shall use the net-pressure only. To simplify notation, we shall omit the subscript '*net*' when using the equations (2.13) – (2.15).

### 3. Complete and basic solutions in near-edge zone. Asymptotic umbrella. Matching constants

We shall call the solution of the problem, formulated in the previous section and accounting for the lag, the *complete solution*. Consider now the simplified basic (outer) problem, in which the lag is neglected. We call this problem outer or *basic problem*, and values referring to it are supplied with the subscript *b*. In the basic problem, the fluid front $L_f$ coincides with the fracture contour $L$, so that there is no need to satisfy the BC (2.14), (2.15). It is assumed that the solution $w_b$, $p_b$ of the basic problem is found analytically or numerically.

*3.1. Equations in a near-edge zone.* Both the complete solution $w$, $p$ and the basic solution $w_b$, $p_b$ meet conditions of plane state for the elasticity in a sufficiently thin strip near the fracture contour (see, e.g. Peirce & Detournay 2008, Appendix B1). The basic solution meets also plane-state fluid equations in a narrow strip (see, e.g. Peirce & Detournay 2008, Linkov 2011, 2012). If the lag is small enough, the plane-state fluid equations are met in the strip for the complete solution, as well. Suppose that it is the case and that at a considered point of the fracture contour, the plane state conditions are met at the distance $r_m$ from the point. In further discussion, the distance $r_m$ will be associated with the matching distance.

Under these assumptions, in the vicinity of the considered point, the elasticity equation (2.13), continuity equation (2.2) and movement equation (2.3) take forms similar for both complete and basic solutions:

$$p(r) = A_R w + \frac{E\prime}{4\pi}\int_0^{r_m}\frac{dw}{d\xi}\frac{d\xi}{r-\xi}, \quad p_b(r) = A_R w_b + \frac{E\prime}{4\pi}\int_0^{r_m}\frac{dw_b}{d\xi}\frac{d\xi}{r-\xi}, \quad 0 \leq r \leq r_m, \quad (3.1)$$

$$\frac{\partial w}{\partial t} + \frac{\partial}{\partial r}[(v_* - v)w] + q_l = 0 \,, \quad \frac{\partial w_b}{\partial t} + \frac{\partial}{\partial r}[(v_* - v_b)w_b] + q_l = 0, \quad 0 \leq r \leq r_m, \quad (3.2)$$

$$v = \left(\frac{1}{\mu\prime}w^{n+1}\frac{dp}{dr}\right)^{\frac{1}{n}}, \qquad v_b = \left(\frac{1}{\mu\prime}w_b^{n+1}\frac{dp_b}{dr}\right)^{\frac{1}{n}}, \qquad 0 \leq r \leq r_m, \quad (3.3)$$

where $A_R$ is the part of the additive operator $A$ outside a vicinity of the point, $E\prime = \frac{E}{1-v^2}$, $E$ is the elasticity modulus, $v$ is the Poisson ratio.

In the second of (3.2), it is assumed that the basic solution provides the propagation speed $v_*$ to an acceptable accuracy. Suppose that at the distance $r_m$ the influence of local perturbation, caused by the lag, is small enough so that the basic solution at this distance equals the actual solution to the acceptable accuracy, as well. Then by subtracting the second of (3.1) from the first we obtain:

$$p(r) = \begin{cases} p_b(r) + \frac{E\prime}{4\pi}\int_0^{r_m}\left(\frac{dw}{d\xi} - \frac{dw_b}{d\xi}\right)\frac{d\xi}{r-\xi}, & 0 \leq r \leq r_m \\ p_b(r), & r \geq r_m \end{cases}. \qquad (3.4)$$

In view of continuity of the actual pressure $p(r)$ at $r = r_m$, equation (3.4) implies:

$$\int_0^{r_m} \left(\frac{dw}{d\xi} - \frac{dw_b}{d\xi}\right)\frac{d\xi}{r-\xi} = 0, \qquad\qquad r = r_m. \qquad (3.5)$$

Equation (3.5) is obtained from the assumption that the influence of local perturbation in the close vicinity of the fracture tip, caused by the lag, becomes negligibly (to an accepted accuracy) small at the distance $r_m$. Therefore, the same equality also holds for $r \ge r_m$. Thus we arrive at the approximate equation:

$$\int_0^{r_m} \left(\frac{dw}{d\xi} - \frac{dw_b}{d\xi}\right)\frac{d\xi}{r-\xi} = 0, \qquad\qquad r \ge r_m. \qquad (3.6)$$

The continuity equations (3.2) yield simplifications, as well. It can be expected, and posterior analysis of the basic solution may serve to verify it, that the convective terms dominate the time derivative within the interval $0 \le r \le r_m$ (cf. Appendix B.2 of the paper by Peirce and Detournay 2008). Then neglecting the time derivative in (3.2) and integrating from the fluid front $r = \lambda$, we obtain:

$$v = v_* + \frac{1}{w(r)}\int_\lambda^r q_l\,dr \;,\; v_b = v_* + \frac{1}{w_b(r)}\int_0^r q_l\,dr, \qquad 0 \le r \le r_m. \qquad (3.7)$$

The integration in the second of (3.7) it is performed from zero, since the basic solution corresponds to zero lag. The leak-off term $q_l$ is now a function of the distance $r - \lambda$ from the front because $t - t_* = (r - \lambda)/v_*$ (for the basic solution, $\lambda = 0$). Thus, for the power-type leak-off term (2.5), we have:

$$q_l = 2C_l v_*^{\beta_l}(r - \lambda)^{-\beta_l}, \qquad\qquad 0 \le \beta_l < 1. \qquad (3.8)$$

Substitution (3.8) into (3.7) and integration yields the forms of the continuity equation near the fracture contour:

$$v_* = v(r) - \frac{2C_l v_*^{\beta_l}}{1-\beta_l}\frac{(r-\lambda)^{1-\beta_l}}{w(r)} \;,\; v_* = v_b(r) - \frac{2C_l v_*^{\beta_l}}{1-\beta_l}\frac{r^{1-\beta_l}}{w_b(r)}, \qquad 0 \le r \le r_m. \qquad (3.9)$$

The equations (3.9) correspond to the steady propagation of a fracture front with the speed $v_*$. The size $r_m$ of their applicability may be established in numerical experiments when finding the basic solution.

*3.2. Asymptotic umbrella. Universal asymptotic solution.* Further simplification involves the assumption that the interval $0 \le r \le r_m$ is under the "asymptotic umbrella". The latter term is incidentally used in the paper by Peirce and Detournay (2008). We adopt it because it nicely suits for the discussion of analytical and computational benefits provided by the asymptotics of the solution. The term means that in the considered near-edge zone, the opening is given by the universal asymptotics $w_a(r)$. As shown in the paper by the author (Linkov 2014), when neglecting the lag, the universal asymptotics may be approximated by the almost monomial equation:

$$w_a(r) = A_w(v_*)r^\alpha, \qquad\qquad 0 \le r < \infty, \qquad (3.10)$$

where $A_w$ is the intensity factor, $\alpha$ is the exponent, characterizing asymptotics. The explicit formulae for $A_w$ and $\alpha$, derived in the cited paper, are reproduced in Appendix. Remarkably, apart from the prescribed properties of the fluid and embedding rock, the both quantities depend merely on the local fracture propagation speed. Moreover, both the intensity factor and the exponent are actually constant in wide ranges of the propagation regimes.

In this study, we are interested in distances far enough from the zone of perturbation caused by the lag. Outside the perturbation zone, we may assume that $A_w$ and $\alpha$ are *constant* up to the distance $r_m$. Then the dependence (3.10) is monomial and the corresponding pressure is (e.g. Spence & Sharp 1985; Garagash et al. 2011):

$$p_a(r) = -A_w(v_*)B(\alpha)E'r^{\alpha-1}, \qquad 0 \leq r < \infty, \qquad (3.11)$$

where

$$B(\alpha) = \frac{\alpha}{4}\cot[\pi(1-\alpha)]. \qquad (3.12)$$

The pair of functions $w_a(r)$ and $p_a(r)$, defined respectively by (3.10) and (3.11), identically satisfy the plane elasticity equation for a semi-infinite crack:

$$p_a(r) = \frac{E'}{4\pi}\int_0^\infty \frac{dw_a}{d\xi}\frac{d\xi}{r-\xi}, \qquad 0 \leq r < \infty. \qquad (3.13)$$

The assumption that the basic solution is under the asymptotic umbrella is expressed by equation:

$$w_b(r) = w_a(r) = A_w(v_*)r^\alpha, \qquad 0 \leq r \leq r_m. \qquad (3.14)$$

Then from the plane elasticity theory (Muskhelishvili, 1975) and the theory of singular integrals (Muskhelishvili, 1953), it follows that the basic net-pressure, corresponding to the basic opening (3.14), is of the form:

$$p_b(r) = p_a(r) + C_m, \qquad 0 \leq r \leq r_m, \qquad (3.15)$$

where $C_m$ is the constant equal to the difference between the basic pressure at the distance $r_m$ and the asymptotic pressure $p_a$ at this distance:

$$C_m = p_b(r_m) - p_a(r_m). \qquad (3.16)$$

Since it has been assumed that for $r \geq r_m$, the basic pressure represents the pressure defined by the complete solution, one can also change $p_b(r_m)$ to $p(r_m)$ on the r. h. s. of (3.16).

*3.3. Matching constants.* We have known both the basic $w_b(r)$, $p_b(r)$ and asymptotic $w_a(r)$, $p_a(r)$ solutions. An analysis of the basic opening $w_b(r)$ provides us with the maximal distance $r_m$, at which $w_b(r)$ can be approximated by the monomial asymptotics (3.14). Then equation (3.16) defines the constant $C_m$. Thus $r_m$ and $C_m$ are known matching constants, which serve us to match the basic (outer) solution with the inner solution to be found in the next section.

When $r_m \to 0$, the constant $C_m$ presents the next (finite) term in the asymptotic expansion of the basic net-pressure. Thus, if having an analytical formula for $p_b(r)$, the limit value of the constant may be found analytically. Such a case is considered in Section 6. Then comparison of the analytical opening $w_b(r)$ with that given by (3.10) provides an estimation of the accuracy of the asymptotic representation (3.14) in the range $0 \leq r \leq r_m$ for various $r_m$. This estimation is also given in Section 6. In the case, when the basic solution $p_b(r)$ is found numerically, say by the explicit (Sethian 1990), or implicit (Peirce & Detournay 2008) level set method, the estimation of $r_m$ and the accuracy may be obtained by comparison of the found basic opening with the asymptotic opening (3.10) at nodal points with increasing distance from the fracture contour.

### 4. Auxiliary problem. Inner solution. Matching condition

*4.1. Auxiliary elasticity equation.* Turn to formulation of the auxiliary (inner) problem. Substitution of (3.14) and (3.15) into the upper line of (3.4) yields:

$$p(r) = C_m + p_a(r) + \frac{E'}{4\pi} \int_0^{r_m} \left( \frac{dw}{d\xi} - \frac{dw_a}{d\xi} \right) \frac{d\xi}{r-\xi}, \qquad 0 \leq r \leq r_m. \quad (4.1)$$

We continuously extend the actual opening $w(r)$ from the strip $0 \leq r \leq r_m$ to the semi-infinite interval $r \geq r_m$ by introducing the auxiliary (inner) opening $w_i(r)$ defined as:

$$w_i(r) = \begin{cases} w(r), & 0 \leq r \leq r_m \\ w_a(r), & r \geq r_m \end{cases}. \quad (4.2)$$

The continuity of $w_i(r)$ at $r = r_m$ follows from the equations $w(r_m) = w_b(r_m) = w_a(r_m)$, assumed to be met to an accepted accuracy.

Since by the definition (4.2), $w_i(r) - w_a(r) = 0$ for $r \geq r_m$, we have for the integral on the r. h. s. of (4.1):

$$\int_0^{r_m} \left( \frac{dw}{d\xi} - \frac{dw_a}{d\xi} \right) \frac{d\xi}{r-\xi} = \int_0^{\infty} \left( \frac{dw_i}{d\xi} - \frac{dw_a}{d\xi} \right) \frac{d\xi}{r-\xi}, \qquad 0 \leq r \leq r_m. \quad (4.3)$$

Using (4.3) in (4.1) and taking into account the identity (3.13), we obtain:

$$p(r) - C_m = \frac{E'}{4\pi} \int_0^{\infty} \frac{dw_i}{d\xi} \frac{d\xi}{r-\xi}, \qquad 0 \leq r \leq r_m. \quad (4.4)$$

Consider the integral on the r. h. s. of (4.4) for $r \geq r_m$. We have by successive using (4.2), (3.6) and (3.14):

$$\int_0^{\infty} \frac{dw_i}{d\xi} \frac{d\xi}{r-\xi} = \int_0^{r_m} \frac{dw}{d\xi} \frac{d\xi}{r-\xi} + \int_{r_m}^{\infty} \frac{dw_a}{d\xi} \frac{d\xi}{r-\xi} = \int_0^{r_m} \frac{dw_b}{d\xi} \frac{d\xi}{r-\xi} + \int_{r_m}^{\infty} \frac{dw_a}{d\xi} \frac{d\xi}{r-\xi} = \int_0^{\infty} \frac{dw_a}{d\xi} \frac{d\xi}{r-\xi}.$$

In view of the identity (3.13), this yields:

$$\frac{E'}{4\pi} \int_0^{\infty} \frac{dw_i}{d\xi} \frac{d\xi}{r-\xi} = p_a(r), \qquad r \geq r_m. \quad (4.5)$$

Combining (4.4) and (4.5) we obtain the *auxiliary elasticity equation* on the entire half-axis $r \geq 0$:

$$\frac{E'}{4\pi} \int_0^{\infty} \frac{dw_i}{d\xi} \frac{d\xi}{r-\xi} = p_i(r), \qquad 0 \leq r < \infty, \quad (4.6)$$

where the inner pressure $p_i(r)$ is defined on the entire half-axis $r \geq 0$ as:

$$p_i(r) = \begin{cases} p(r) - C_m, & 0 \leq r \leq r_m \\ p_a(r), & r \geq r_m \end{cases}. \tag{4.7}$$

The inner pressure $p_i(r)$ is continuous at the point $r_m$ by the definition (3.16) of the matching constant $C_m$ and by the assumption $p(r_m) = p_b(r_m)$.

*4.2. Auxiliary fluid equation and conditions in lag-zone.* In view of (4.2) and (4.7), for $0 \leq r \leq r_m$, the auxiliary elasticity equation (4.6) involves as unknowns the actual net pressure $p(r)$ and the actual opening $w(r)$. For them we have the additional dependences, given by the first equations in (3.3) and (3.7) in the region of fluid flow, the boundary conditions (2.14) and (2.15) at the lag area, and the fracture condition (2.8), which defines the lag $\lambda$. For convenience, we re-write them in terms of the auxiliary quantities as:

$$\left(\frac{1}{\mu'} w_i^{n+1} \frac{dp_i}{dr}\right)^{\frac{1}{n}} = v_* + \frac{2C_l v_*^{\beta_l}(r-\lambda)^{1-\beta_l}}{w_i(r)(1-\beta_l)}, \qquad \lambda \leq r \leq r_m, \tag{4.8}$$

$$p_i(r) = -(\sigma_0 + C_m), \qquad 0 \leq r \leq \lambda, \tag{4.9}$$

$$w_i(r) = \sqrt{\frac{2}{\pi}} \frac{4}{E'} K_{IC} r^{\frac{1}{2}} + O\left(r^{\frac{3}{2}}\right). \tag{4.10}$$

From the elasticity theory (see, e. g. Muskhelishvili 1975) it follows that it is impossible to simultaneously prescribe both the displacements and the tractions at any part of the boundary. Meanwhile, according to the second lines in (4.2) and (4.7), both the auxiliary opening $w_i(r)$ and the auxiliary pressure $p_i(r)$ are assumed prescribed for $r > r_m$. This is the consequence of writing the symbols of exact equality in equations involving the assumption that the complete and basic solutions practically coincide for $r \geq r_m$. We remove the artificial, in essence, difficulty and simplify the problem by extending the upper boundary of the interval for equation (4.8) to infinity:

$$\left(\frac{1}{\mu'} w_i^{n+1} \frac{dp_i}{dr}\right)^{\frac{1}{n}} = v_* + \frac{2C_l v_*^{\beta_l}(r-\lambda)^{1-\beta_l}}{w_i(r)(1-\beta_l)}, \qquad \lambda \leq r < \infty. \tag{4.11}$$

This excludes the distance $r_m$, the asymptotic opening $w_a(r)$ and the asymptotic pressure $p_a(r)$ from the formulation of the auxiliary (inner) problem (4.6), (4.9) - (4.11). The asymptotics $w_a(r)$ is used *after* solving the auxiliary problem to verify the possibility to change the finite interval in (4.8) to the semi-infinite interval in (4.11). It serves to formulate the matching condition.

*4.3. Auxiliary system in normalized variables.* The auxiliary problem (4.6), (4.9) - (4.11) corresponds to the steady propagation of a semi-infinite hydraulic fracture with the speed $v_*$ and the lag $\lambda$ in host rock with the normal pressure $\sigma_0' = \sigma_0 + C_m$. The problem looks mathematically consistent. In the particular case of a Newtonian fluid ($n = 1$) and viscosity dominated regime ($C_l = 0$), its solution has been given by Garagash and Detournay (2000). The only difference is that since the matching problem was beyond the scope of these authors, they did not include the matching constant $C_m$ in the r. h. s. of the boundary condition (4.9). This difference, although crucial for the method developed, is insignificant in the computational sense when solving the auxiliary system.

It is convenient to reformulate the problem by using dimensionless variables and parameters:

$$\xi = \frac{r}{L_\mu}, \; \Omega_i = \frac{w_{aux}}{L_\mu \varepsilon}, \; \Pi_i = \frac{p_{aux}}{\sigma_0'}, \; k = \left(\frac{L_k}{L_\mu}\right)^{1/2}, \; \varepsilon = \frac{\sigma_0'}{E'}, \; \psi = \frac{2C_l}{\varepsilon L_\mu^{\beta_l} v_*^{1-\beta_l}(1-\beta_l)},$$

$$L_\mu = \left(\frac{E'}{\sigma_0'}\right)^{\frac{n+2}{n}} t_n v_*, \; L_k = \frac{32}{\pi}\left(\frac{K_{IC}}{\sigma_0'}\right)^2, \; \sigma_0' = \sigma_0 + C_m, \; t_n = \left(\frac{\mu'}{E'}\right)^{\frac{1}{n}}. \qquad (4.12)$$

Then the auxiliary problem becomes:

$$\frac{1}{4\pi}\int_0^\infty \frac{d\Omega_i}{d\eta}\frac{d\eta}{\xi-\eta} = \Pi_i(\xi), \qquad\qquad 0 \le \xi < \infty, \qquad (4.13)$$

$$\left(\Omega_i^{n+1}\frac{d\Pi_i}{d\xi}\right)^{\frac{1}{n}} = 1 + \psi\frac{(\xi-\Lambda)^{1-\beta_l}}{\Omega_i(\xi)}, \qquad \Lambda \le \xi < \infty, \qquad (4.14)$$

$$\Pi_i = -1, \qquad\qquad\qquad 0 \le \xi \le \Lambda, \qquad (4.15)$$

$$\Omega_i(\xi) = k\xi^{\frac{1}{2}} + O\left(\xi^{\frac{3}{2}}\right), \qquad\qquad \xi \to 0. \qquad (4.16)$$

The normalized lag $\Lambda$ is unknown and it is found from the condition (4.16). As noted by Garagash and Detournay (2000), to avoid iterations in $\Lambda$, when satisfying this condition, it is reasonable to prescribe $\Lambda$ and to find the corresponding $k$. In general, when leak-off is not neglected, the problem (4.13) - (4.16) contains three parameters: $k$, $\beta_l$ and $\psi$. For non-zero leak-off ($\psi \ne 0$), the parameter $\psi$ may be excluded by re-normalizing variables (see Appendix). In the case of zero leak-off ($\psi = 0$), there is the only parameter $k$.

For zero value of the lag ($\Lambda = 0$), the system turns into the universal system (A1) – (A3), considered in Appendix. It corresponds to the problem of steady (with the speed $v_*$) propagation of a semi-infinite fracture driven by a power-law fluid with power-law leak-off when the fluid front coincides with the fracture tip. In this case, the boundary condition on the lag surface drops out from the system, and the normalizing value of $\sigma_0'$, used in (4.12), may be taken as $E'$. Then $L_\mu = t_n v_*$. With these normalizing values, the universal almost monomial asymptotic solution is given in Appendix.

*4.4. Matching condition.* The influence of the local perturbation, caused by the lag, may be evaluated by the comparison of the solution of the auxiliary (inner) problem (4.13) - (4.16) accounting for the lag, with the universal solution of the analogous problem (A1) - (A3) neglecting the lag. The comparison provides the *minimal* normalized distance $\xi_a$, at which the basic solution agrees with the inner solution to an accepted accuracy. (Naturally, the asymptotic solution should be renormalized to the same variables, which are used in (4.13) - (4.16)). When having $\xi_a$, the corresponding dimensional value $r_a = \xi_a L_\mu$ gives the *minimal distance*, at which the auxiliary opening $w_i(r)$ agrees (to the accepted accuracy) with the assumed universal asymptotics (3.10) of the basic opening $w_b(r)$. In contrast with the matching distance $r_m$, defined by the *outer* (basic) solution, the distance $r_a$ is defined by the *inner* solution, while in each of the cases, the universal asymptotic solution serves for the comparison.

Clearly, the matching of the basic and inner solutions is possible only when

$$r_m \geq r_a. \tag{4.17}$$

The inequality (4.17) presents the *matching condition*. It is to be verified after solving the auxiliary problem. Matching to the accepted accuracy is impossible if the condition (4.17) is not satisfied. In such cases, it might be reasonable to decrease the accepted level of accuracy.

### 5. Complete solution

When the condition (4.17) is fulfilled, the complete solution, accounting for the lag, is:

$$w(r) = \begin{cases} w_b(r) & r \geq r_m \\ w_i(r) & 0 \leq r \leq r_m \end{cases}, \quad p(r) = \begin{cases} p_b(r) & r \geq r_m \\ p_i(r) + C_m & 0 \leq r \leq r_m \end{cases}. \tag{5.1}$$

The length of the lag $\lambda$ is known from the inner solution.

Summarizing, the method accounting for the lag comprises of the next steps.

I. The basic solution $w_b(r)$, $p_b(r)$ to a problem for a hydraulic fracture, propagating, in general, in 3D, is found neglecting the lag. When finding it numerically for a small lag, the size of a spatial grid may be notably greater than the lag zone. Meanwhile it should be less than distance from the fracture contour covered by the asymptotic umbrella. The solution provides also the fracture *propagation speed* $v_*$ at points of the fracture front. For a considered point at a considered time, the speed $v_*$ defines the universal asymptotic opening $w_a(r)$ and the corresponding asymptotic pressure $p_a(r)$. They are employed to evaluate (i) the *maximal matching distance* $r_m$, at which the asymptotics $w_a(r)$ is yet applicable to an accepted accuracy, and (ii) the *matching pressure constant* $C_m = p_b(r_m) - p_a(r_m)$ at the matching distance. The universal asymptotics is found by using equations given in Appendix.

II. The 1D auxiliary problem (4.13) - (4.16) at the half-axis $[0, \infty)$ is solved numerically with the normalizing values defined by (4.12), where the propagation speed $v_*$ and the normalizing pressure $\sigma_0' = \sigma_0 + C_m$ are known from the basic solution. The solution defines the normalized lag $\Lambda$, the normalized inner opening $\Omega_i(\xi)$ and the normalized auxiliary pressure $\Pi_i(\xi)$. It also defines the normalized *minimal distance* $\xi_a$, at which the inner opening $\Omega_i(\xi)$ practically coincides with similarly normalized asymptotic opening $\Omega_a(\xi)$. The definitions of the normalized values (4.12) provide the corresponding dimensional lag $\lambda$, the inner opening $w_i$, the auxiliary pressure $p_i$ and the dimensional *minimal matching distance* $r_a$.

III. The matching condition $r_m \geq r_a$ is checked. If it is met, the matching is justified. Otherwise, matching is impossible. Then it might be reasonable to decrease the level of accepted accuracy.

IV. When the matching condition is met, the solution of the problem accounting for the lag, is found by using the basic and auxiliary solutions in (5.1).

## 6. Example: solution of Spence & Sharp problem accounting for the lag

Apply the method developed to the problem by Spence & Sharp (1985). We shall account for the lag. The problem consists in finding the pressure and the opening as functions of the time $t$ and the spatial coordinate $x$ under the conditions of plane-strain and plane flow in the plane $(x, y)$ orthogonal to the fluid front $|x| = x_*$ and the fracture contour $|x| = x_0$; thus the lag is $\lambda = x_0 - x_*$. The fluid is Newtonian ($n = 1$) with the dynamic viscosity $M = \mu$ ($\mu' = 12\mu$). The host rock has the Young modulus $E$ and the Poisson ratio $\nu$ ($E' = \frac{E}{1-\nu^2}$). The formation is assumed impermeable ($C_l = 0$). The injection rate at the inlet $x = 0$ per unit length is $Q_0$. The fracture propagates symmetrically about the origin. In the considered problem, the fluid equations (2.2), (2.3) and the elasticity equation (2.6) are, respectively:

$$\frac{\partial w}{\partial t} + \frac{\partial (wv)}{\partial x} = 0, \qquad\qquad |x| \leq x_*, \qquad (6.1)$$

$$v = -\frac{w^2}{\mu'}\frac{\partial p}{\partial x}, \qquad\qquad |x| \leq x_*, \qquad (6.2)$$

$$p(x) = -\frac{E'}{4}\int_{-x_0}^{x_0}\frac{dw}{d\xi}\frac{d\xi}{x-\xi}, \qquad\qquad |x| \leq x_0. \qquad (6.3)$$

(Recall that we have agreed to denote $p(x)$ the net-pressure). The BC at the inlet is:

$$(wv)^+ = -(wv)^- = \frac{Q_0}{2}, \qquad\qquad x = 0. \qquad (6.4)$$

At the fracture contour, the condition of zero opening (2.7) becomes:

$$w(|x_0|, t) = 0. \qquad (6.5)$$

At the fracture tip, the condition (2.8) is met. In the form (4.10) it reads (see, e. g. Rice 1968):

$$w(r) = \sqrt{\frac{2}{\pi}}\frac{4}{E'}K_{IC}r^{\frac{1}{2}} + O\left(r^{\frac{3}{2}}\right), \qquad (6.6)$$

where $r = x_0 - |x|$.

At the lag, the BC (2.14), (2.15) become:

$$p(x, t) = -\sigma_0, \qquad\qquad x_* \leq |x| \leq x_0. \qquad (6.7)$$

The propagation of the fluid front is controlled by the speed equation (2.5), which now reads:

$$v_* = \frac{dx_*}{dt} = -\frac{w^2}{\mu'}\frac{\partial p}{\partial x}\Big|_{|x|=x_*}. \qquad (6.8)$$

The IC (2.11), (2.12) at the initial time $t_0 = 0$ are, respectively:

$$w(x, 0) = 0, \qquad (6.9)$$

$$x_*(0) = x_0(0) = 0. \qquad (6.10)$$

We need to solve the system of equations $(6.1) - (6.3)$ under the BC $(6.4) - (6.7)$, the SE $(6.8)$ and the IC $(6.9)$, $(6.10)$. In accordance with the method suggested, we shall match the basic solution with the solution of the auxiliary problem.

*Step I. Solution of the basic (outer) problem.* The basic solution corresponds to neglecting the lag ($\lambda = 0$,) and fracture toughness ($K_{Ic} = 0$). Then in $(6.1)$, $(6.2)$ and $(6.8)$ we have $x_* = x_0$, while equation $(6.7)$ drops out from the system. The solution was found by Adachi and Detournay (2002) numerically and in the paper by the author (Linkov 2012) analytically. For our purpose, it is convenient to use the analytical solution. The solution is self-similar in the variables $\zeta$, $\xi_*$, $\psi(\zeta)$, $P(\zeta)$, $V(\zeta)$, $V_*$, connected with the physical quantities by equations:

$$\zeta = x/x_*, \ x_* = x_n \xi_*(t/t_n)^{2/3}, \ w_b = w_n \xi_*(t/t_n)^{1/3}\psi(\zeta), \ p_b = E'(t/t_n)^{-1/3}P(\zeta), \qquad (6.11)$$

$$v_b = v_n \xi_*(t/t_n)^{-1/3}V(\zeta), \ v_* = \frac{2}{3}v_n \xi_*(t/t_n)^{-1/3},$$

where the normalizing quantities are defined as: $t_n = \mu'/E'$, $x_n = w_n(Q_0 t_n)^{1/2}$, $v_n = Q_0/w_n$. The solution is (Linkov 2012): $\xi_* = 0.6157$, $V(1) = 2/3$, $\psi(\varsigma) = 2.7494(1-\varsigma)^{2/3}[1 - 0.3040(1-\varsigma) - 0.0231(1-\varsigma)^2]$, $P(\varsigma) = -0.2188[P_0(\varsigma) - 0.4965P_1(\varsigma) + 0.06031P_2(\varsigma)] + 0.0625(2 - \pi\varsigma)$, $V(\varsigma) = \frac{2}{3}[1 - 0.1(1-\varsigma) + 0.1026(1-\varsigma)^2]$, where $P_0(\varsigma) = J_0(1-\varsigma) + J_0(1+\varsigma)$, $P_1(\varsigma) = J_1(1-\varsigma) + J_1(1+\varsigma)$, $P_2(\varsigma) = J_2(1-\varsigma) + J_2(1+\varsigma)$ with $J_0(z) = 2/3z^{-1/3}f(z) - 1/(1-z)$, $J_1(z) = S_0(z) + zJ_0(z)$, $J_2(z) = S_1(z) + zJ_1(z)$, $S_0(z) = 3/2 + z^{2/3}f(z)$, $S_1(z) = 3/5 + zS_0(z)$, $f(z) = \ln|1 - z^{1/3}| - 1/2\ln|1 + z^{1/3} + z^{2/3}| + \sqrt{3}\{-\pi/6 + \text{atan}[(1 + 2/z^{1/3})/\sqrt{3}]\}$.

Expansion of the solution near the fracture tip ($\zeta = 1$) yields the asymptotic expressions for the self-similar opening $\psi(\zeta)$ and net-pressure $P(\zeta)$: $\psi(\zeta) = \psi_a\left(\frac{r}{x_*}\right)$, $P(\zeta) = P_a\left(\frac{r}{x_*}\right) + C_{P0}$, where $\psi_a\left(\frac{r}{x_*}\right) = A_\psi \left(\frac{r}{x_*}\right)^{\frac{2}{3}}$, $P_a\left(\frac{r}{x_*}\right) = A_P\left(\frac{r}{x_*}\right)^{-\frac{1}{3}}$ with $A_\psi = 2^{\frac{2}{3}}\sqrt{3}$, $A_P = \frac{1}{3\sqrt[3]{2}}$, $C_{P0} = 0.68793$. As it should be, the corresponding non-normalized quantities $w_a(r)$, $p_a(r)$, evaluated by using these asymptotic expressions, coincide with those following from the universal asymptotics given in Appendix. Specifically, in the considered case of a Newtonian fluid and zero leak-off, we have the viscosity dominated regime, for which equations (A11) of Appendix yield the monomial solution $(3.10)$, $(3.11)$ with $\alpha = \frac{2}{3}$, $A_w(v_*) = \frac{36^{2/3}}{2\sqrt{3}}(v_* t_n)^{\frac{1}{3}}$, $B(\alpha) = \frac{1}{6\sqrt{3}}$, where the time constant is $t_n = \frac{\mu'}{E'}$. Therefore

$$w_a(r) = \frac{36^{2/3}}{2\sqrt{3}}(v_* t_n)^{\frac{1}{3}} r^{\frac{2}{3}}, \ p_a(r) = -\frac{1}{36^{1/3}} E'(v_* t_n)^{\frac{1}{3}} r^{-\frac{1}{3}}. \tag{6.12}$$

Now we need to compare the basic and asymptotic solutions to find the maximal matching distance $r_m$, at which the approximations (3.14), (3.15), defined by the universal asymptotics (6.12), are applicable to an accepted accuracy. Then the matching pressure constant $C_m$ becomes known as the difference $C_m = p_b(r_m) - p_a(r_m)$. The comparison is presented in Fig. 1 and Fig. 2 in self-similar variables.

Fig. 1 shows the relative difference between the basic and asymptotic opening on the entire fracture. It can be seen that the accuracy of the approximation (6.12) is 5.85% when $r_m = 0.2x_*$ ($x/x_* = 0.8$); it is 2.54% when $r_m = 0.1x_*$ ($x/x_* = 0.9$). Therefore the approximation (3.14) may serve for matching of the opening to the accuracy of 2.54% with $r_m = 0.1x_*$.

Fig. 2 shows the difference $C_P$ between the self-similar basic $P(\zeta)$ and self-similar asymptotic $P_a(\zeta)$ pressure. The difference $C_P$ defines the matching constant $C_m$ (3.16) in non-normalized variables: $C_m = C_P E'(t/t_n)^{-\frac{1}{3}}$. Remarkably, it is almost constant: the limiting value $C_P(1) = 0.688$, when approaching to the fracture tip ($x \to x_*$), differs from the value $C_P(0) = 0.791$ at the inlet ($x = 0$) merely 15%. The self-similar matching constant $C_P$ equals 0.755, 0.739, 0.726, when the distance $r_m$ from the fracture tip is $0.2x_*$, $0.1x_*$, $0.05x_*$, respectively. To the accuracy of $C_m$, not less than 3.7 %, we may take the mean value $C_P = 0.714$ in the interval $0 \le r_m \le 0.1x_*$. Therefore, the approximation (3.15) may serve for matching the net-pressure to the accuracy of 3.7% with $C_m = 0.714 E'(t/t_n)^{-1/3}$, $r_m = 0.1x_*$.

From the comparison, it follows that with a relative error not exceeding four percent, we may use the matching distance $r_m = 0.1x_*$ and the matching pressure $C_m = 0.714 E'(t/t_n)^{-1/3}$

*Step II. Solution of auxiliary (inner) problem.* Having $\alpha$, $A_w$, $v_*$, $r_m$ and $C_m$ from the basic solution, we start the step II. In the considered case ($n = 1$, $\psi = 0$), the auxiliary problem (4.13) - (4.16) is solved by Garagash and Detournay (2000). (Recall that we need to normalize the pressure by $\sigma_0' = \sigma_0 + C_m$ instead of normalization by $\sigma_0$ used in the cited paper). The detailed data on the normalized lag $\Lambda$, the normalized inner opening $\Omega_i(\xi)$, and the normalized auxiliary net-pressure $\Pi_i(\xi)$ are given in Fig. $3 - 9$ of the paper by Garagash and Detournay (2000) (in the paper by these authors, $\Omega_i(\xi)$ and $\Pi_i(\xi)$ are denoted as $\Omega_\infty(\xi)$ and $\Pi_\infty(\xi)$, respectively). We refer a reader to their comprehensive results.

Having the inner solution, we focus on the distance $r_a$, at which the influence of the perturbation zone is negligible. The distance $r_a$ may be evaluated from the graphs on Fig. 4 - 6 of the paper by Garagash and Detournay (2000). Unfortunately, the graphs do not allow us to have the distance to the accuracy better than 5 % even when magnifying these graphs three-fold. To this accuracy, for zero critical SIF ($K_{IC} = 0$), the asymptotic equations (6.12) are met at the normalized distance $\xi_a = \frac{r_a}{L_\mu} = 6$. Thus the minimal distance, at which the influence of the perturbation caused by the lag may be neglected to the accuracy of five percent, is $r_a = L_\mu \xi_a = 6L_\mu$.

*Step III. Checking the matching condition.* With $r_a = L_\mu \xi_a$, the matching condition can be written as:

$$\frac{x_*}{L_\mu} \geq \frac{\xi_a}{r_m/x_*} . \tag{6.13}$$

After using the definitions $L_\mu = (E'/\sigma_0')^3 v_* t_n$, $\sigma_0' = \sigma_0 + C_m$, $C_m = C_P E'(t/t_n)^{-1/3}$, the condition (6.13) reads:

$$\frac{x_*}{v_*}\left[\frac{\sigma_0'}{E'} + C_P\left(\frac{t}{t_n}\right)^{-\frac{1}{3}}\right]^3 \geq \frac{\xi_a}{r_m/x_*} . \tag{6.14}$$

The basic solution implies that the ratio $\frac{x_*}{v_*}$ in (6.14) does not depend on the pumping rate $Q_0$: $\frac{x_*}{v_*} = \frac{3}{2}t$. Hence, the matching condition (6.14) may be written in terms of the time:

$$t \geq t_m , \tag{6.15}$$

where

$$t_m = t_n\left(\frac{E'}{\sigma_0'}\right)^3\left[\left(\frac{2}{3}\frac{\xi_\infty}{r_m/x_*}\right)^{\frac{1}{3}} - C_P\right]^3 \tag{6.16}$$

is the acceptable matching time. The outer (basic) and the inner (auxiliary) solutions may be matched when $t \geq t_m$. Notably, in the problem considered, the matching condition (6.15) does not depend on the pumping rate $Q_0$.

Commonly the constant $C_P$ on the r. h. s. of (6.16) is much less than the first term in brackets. Neglecting it, we have the equation simplified:

$$t_m = \frac{2}{3}\frac{\xi_\infty}{r_m/x_*}t_n\left(\frac{E'}{\sigma_0'}\right)^3 . \tag{6.17}$$

Estimate the matching time (6.17) by using the same values $E' = 3\cdot10^4$ MPa, $\sigma_0' = 10$ MPa, $\mu = 10^{-7}$ MPa·s, ($\mu' = 1.2\cdot10^{-6}$ MPa·s) as those used in the paper by Garagash and Detournay (2000). Then $t_n\left(\frac{E'}{\sigma_0'}\right)^3 = 1.08$ s and for the estimated values $r_m/x_* = 0.1$ and $\xi_\infty = 6$, the matching time is $t_m = 43.2$ s. As it is less than minute, the matching condition (6.15) is met for the time range of practical interest.

*Step IV. Complete solution of the problem accounting for the lag.* For $t \geq t_m$, the opening and the net-pressure are defined by equations (5.1). Therein, $w_b(r,t)$, $p_b(r,t)$, $v_*(t)$, $x_*(t)$ are known from the presented self-similar solution and the definitions (6.11); the functions $w_i(r)$, $p_i(r)$ and the lag $\lambda$ are defined by (4.12), where $\Omega_i(\xi)$, $\Pi_i(\xi)$ and the normalized lag $\Lambda$ are given in Fig. 4 - 6 of the paper by Garagash and Detournay (2000). In particular, for the lag, by using the dependence $\frac{x_*}{v_*} = \frac{3}{2} t$, we obtain:

$$\frac{\lambda}{x_*} = \frac{2}{3} \left(\frac{E'}{\sigma_0'}\right)^3 \frac{t_n}{t} \Lambda. \tag{6.18}$$

As shown in the cited paper (p. 187), the normalized lag is maximal for zero critical SIF ($K_{IC} = 0$). In this case, it is $\Lambda = 0.3574$. Then at $t = t_m$, and equation (6.18) gives $\lambda = 0.006 x_*$. The lag is within the range, for which meshes, used in the algorithms suggested for shallow fractures, are applicable (almost 1000 mesh points for a fracture wing (Gordeli & Detournay 2011). This shows that the method developed may serve in the intermediate range between deep and shallow fractures.

## 7. Conclusions

The conclusions of the paper are summarized as follows.

(i) The method suggested provides the solution of 3D hydraulic fracture problem accounting for a zone of local perturbation near the fracture contour by combining the basic solution, which does not account for the perturbation, with the solution of the formulated 1D auxiliary problem.

(ii) The basic solution should account for the leading asymptotics of the opening. The power and the intensity factor of the leading asymptotics are defined by the universal asymptotics, depending on the mere characteristic of a particular flow, the local speed of the fracture propagation. The explicit equations for the exponent and intensity of asymptotic values are given in Appendix. The basic solution serves also to find the *maximal distance* $r_m$ from the fracture contour, at which the asymptotic umbrella is still effective, and the corresponding matching constant $C_m$ for the net-pressure at this distance.

(iii) The inner solution is found by solving 1D problem for steady propagation of a semi-infinite fracture with the speed, equal to that defined by the basic solution, and with the corrected value $-\sigma_0' = -(\sigma_0 + C_m)$ of the net-pressure in the lag-zone. The inner solution serves also to find the *minimal distance* $r_a$ from the fracture contour, at which the influence of the local perturbation, caused by the lag, is negligible to an accepted accuracy. The matching of outer and inner solutions is possible under the condition that $r_m \geq r_a$.

(iv) For the Spence & Sharp (1985) problem, the matching is commonly applicable for time greater than one minute independently on the injection rate. The solution shows that the method developed for deep fractures may also serve in the intermediate range between deep and shallow fractures.

**Acknowledgement.** The author appreciates the support of the EU Marie Curie IAPP program (Grant # 251475).

## Appendix. Universal umbrella

*A1. Universal system in dimensionless variables.* Consider the system of equations of the auxiliary problem in the case of zero lag. Then we arrive at the universal asymptotic system (Linkov 2014). In the normalized values and parameters (4.12), the system for the asymptotic umbrella is:

$$\frac{1}{4\pi}\int_0^\infty \frac{d\Omega_a}{d\eta}\frac{d\eta}{\xi-\eta} = \Pi_a(\xi), \qquad\qquad 0 \le \xi < \infty, \qquad \text{(A1)}$$

$$\left(\Omega_a{}^{n+1}\frac{d\Pi}{d\xi}\right)^{\frac{1}{n}} = 1 + \psi\frac{\xi^{1-\beta_l}}{\Omega_a(\xi)}, \qquad\qquad 0 \le \xi < \infty, \qquad \text{(A2)}$$

$$\Omega_a(\xi) = k\xi^{\frac{1}{2}} + O\left(\xi^{\frac{3}{2}}\right), \qquad\qquad \xi \to 0. \qquad \text{(A3)}$$

It contains two parameters only: the leak-off parameter $\psi$ and the toughness parameter $k$. Since equation (A2) contains also the term 1, there are three limiting cases, discussed below. The parameter $\psi$ may be excluded by renormalizing the opening, net-pressure, coordinate and the coefficient $k$ as:

$$\Omega_1 = \frac{\Omega_a}{\psi^{\alpha_\mu/\varepsilon_\mu}}, \Pi_1 = \Pi_a\psi^{(1-\alpha_\mu)/\varepsilon_\mu}, \xi_1 = \frac{\xi}{\psi^{1/\varepsilon_\mu}}, \ k_1 = \frac{k}{\psi^{(\alpha_\mu-1/2)/\varepsilon_\mu}}, \qquad \text{(A4)}$$

with

$$\alpha_\mu = \frac{2}{n+2}, \ \varepsilon_\mu = \alpha_\mu - (1-\beta_l). \qquad \text{(A5)}$$

Then the system becomes:

$$\frac{1}{4\pi}\int_0^\infty \frac{d\Omega_1}{d\eta_1}\frac{d\eta_1}{\xi_1-\eta_1} = \Pi_1(\xi_1), \qquad\qquad \text{(A6)}$$

$$\left(\Omega^{n+1}\frac{d\Pi}{d\xi_1}\right)^{\frac{1}{n}} = 1 + \frac{\xi_1{}^{1-\beta_l}}{\Omega(\xi_1)}, \qquad\qquad \text{(A7)}$$

$$\lim_{\xi_1\to 0}\frac{\Omega_1(\xi_1)}{\xi_1{}^{1/2}} = k_1. \qquad\qquad \text{(A8)}$$

Actually equations (A6) - (A8) are less convenient than (A1) - (A3), because the parameter $\varepsilon_\mu$, defined in (A5), is commonly quite small what drastically extends the scales of $\xi_1$ and $\Omega_1$. Besides, as clear from (A4), they are inapplicable to the case of negligible leak-off, for which $\psi = 0$. For these reasons, we employ them only for intermediate calculations.

*A2. Limiting regimes.* There are three limiting cases, for which the exact solution is given by the monomial opening (3.10) and the corresponding pressure (3.11); the pair identically satisfies equation (3.13). In terms of the normalized values, the solution is:

$$\Omega_a(\xi) = A\xi^\alpha, \ \Pi_a(\xi) = -AB(\alpha)\xi^{\alpha-1}, \qquad\qquad \text{(A9)}$$

where $B(\alpha)$ is defined by (3.12). The exponent $\alpha$ and the coefficient $A$ depend on a dominating factor of fracture propagation.

(i) *Toughness dominated regime* occurs when the asymptotics of the opening is defined by equation (A3) only. Hence, in (A9):

$$\alpha = \alpha_k = \frac{1}{2}, A = A_k = k, \qquad\qquad \text{(A10)}$$

and the value $L_\mu = 1$ is used in the normalizing quantities (4.12).

(ii) *Viscosity dominated regime* occurs when toughness and leak-off are negligible ($k = 0$, $\psi = 0$). In this case,

$$\alpha = \alpha_\mu = \frac{2}{n+2}, A = A_\mu = [(1-\alpha)B(\alpha)]^{-\frac{1}{n+2}}. \qquad (A11)$$

For a Newtonian fluid ($n = 1$), equations (A11) give the Spence and Sharp (1985) exponent $\alpha = \frac{2}{3}$, while $A_\mu = 2^{1/3}3^{5/6} = 3.147345$.

(iii) *Leak-off dominated regime* occurs when toughness is negligible ($k = 0$) while the second term on the r. h. s. of (A2) is much greater than 1. Then:

$$\alpha = \alpha_l = \frac{n(1-\beta_l)+2}{2n+2}, A = A_l = [(1-\alpha)B(\alpha)\psi^n]^{-\frac{1}{2n+2}}. \qquad (A12)$$

For Carter's leak-off ($\beta_l = 1/2$), equations (A12) give the solution by Lenoach (1985); if the fluid is Newtonian ($n = 1$), equations (A12) yield $\alpha_l = \frac{5}{8}$, $A_l = 2.533559\sqrt[4]{\psi}$.

*A3. Asymptotics for small and large $\xi$. Exact monomial solution for a particular case.* In the general case, the least of the three numbers $\alpha_k$, $\alpha_\mu$ and $\alpha_l$ defines the asymptotics of the solution of (A1) - (A3) when $\xi \to 0$, the largest when $\xi \to \infty$. Therefore, we need to define the mutual positions of these numbers on the real axis.

Since by (A10) $\alpha_k = \frac{1}{2}$, we have $\alpha_\mu > \alpha_k$ for thinning fluids ($0 < n < 1$) and even for thickening fluids with $1 < n < 2$. For thinning fluids, the number $\alpha_l$, as follows from its definition (3.18), is also greater than $\alpha_k$ in cases of practical significance, for which $\beta_l < 1/n$. For the difference $\alpha_\mu - \alpha_l$, we have:

$$\Delta\alpha = \alpha_\mu - \alpha_l = \frac{n}{2n+2}[\alpha_\mu - (1-\beta_l)]. \qquad (A13)$$

By (A11) and (A13), the number $\alpha_l$ is less than $\alpha_\mu$, when $\frac{2}{n+2} > 1 - \beta_l$; it is greater than $\alpha_\mu$, when $\frac{2}{n+2} < 1 - \beta_l$; and the numbers are equal when $\frac{2}{n+2} = 1 - \beta_l$. In the last case, the system (A1) - (A3) has the exact monomial solution (A9) for any $\psi$, when $k = 0$ ($K_{Ic} = 0$), with $\alpha = \alpha_\mu = \alpha_l = \frac{2}{n+2}$ and $A = A_l$ defined by the algebraic equation:

$$A_l^{\frac{n+2}{n}}[(1-\alpha_\mu)B(\alpha_\mu)]^{\frac{1}{n}} = 1 + \frac{\psi}{A_l}.$$

From it, $A_l$ is easily obtained either by plotting the function $\psi(A_l) = A_l^{\frac{2n+2}{n}}[(1-\alpha_\mu)B(\alpha_\mu)]^{\frac{1}{n}} - A_l$, or by proper re-normalizing variables. The exact solution may serve as a benchmark when considering the case $K_{Ic} = 0$ and solving numerically the system (A1), (A2).

*A4. Almost monomial solution for intermediate regimes between toughness dominated and viscosity dominated regimes.* The results for this case are summarized in the form of the "universal asymptote" as a curve in log-log coordinates in Fig. 2 of the paper by Gordeliy and Peirce (2013). To simplify its employing, the analytical approximation of the curve in the piece-wise monomial form may be used:

$$\Omega_{k\mu}(\xi_{k\mu}) = A_{k\mu}\xi_{k\mu}{}^\alpha, \qquad (A14)$$

where now the opening $w$ and the coordinate $r$ are renormalized as $\Omega_{k\mu} = \frac{w}{w_{k\mu}}$, $\xi_{k\mu} = \frac{r}{L_{k\mu}}$ with $w_{k\mu} = \frac{L_k^2}{L_\mu}$, $L_{k\mu} = \frac{L_k^3}{L_\mu^2}$ and $L_k$ and $L_\mu$ are defined in (4.12). The exponent $\alpha$ and the factor $A_{k\mu}$ are almost constant in wide ranges of the normalized coordinate $\xi_{k\mu}$:

$$\alpha = \begin{cases} 1/2 \\ 0.599 \\ 2/3 \end{cases}, \quad A_{k\mu} = \begin{cases} 1 \\ 2^{1/3}3^{5/6} \\ 2^{1/3}3^{5/6} \end{cases} \quad \text{for } \xi_{k\mu} = \begin{cases} \xi_{k\mu} \leq 10^{-5} \\ 10^{-5} \leq \xi_{k\mu} \leq 1. \\ \xi_{k\mu} \geq 1 \end{cases} \tag{A15}$$

The approximations (A15) are obtained for a Newtonian fluid. Their extension to an arbitrary power-law fluid is as follows. For the exponent $\alpha$, the intermediate number 0.599 is close to the mean of the toughness $\alpha_k = \frac{1}{2}$ and viscosity $\alpha_\mu = \frac{2}{n+2}$ exponents. Therefore, in the general case we may take the intermediate value as $\alpha = \frac{n+6}{4n+8}$. Prescribing $A_{k\mu}$ is even simpler: since the values in the last two lines for $A_{k\mu}$ equal to $A_\mu$ for a Newtonian fluid, we may set them as $A_\mu$, defined in (A11) in the general case. Then the same piece-wise monomial equation (A14) with the same normalized variables becomes available for a non-Newtonian fluid.

*A5. Almost monomial solution for intermediate regimes between viscosity dominated and leak off dominated regimes.* Consider the case important for practice, when the rock toughness may be neglected ($k = 0$). In this case, in intermediate calculations, we exclude the parameter $\psi$ by using the renormalized values (A4), (A5). The almost monomial solution in terms of the formerly used variables $\Omega_a$ and $\xi$ is:

$$\Omega_a(\xi) = A_{\mu l}\xi^\alpha, \tag{A16}$$

where $\alpha$ and $A_{\mu l}$ are piece-wise constant functions of the renormalized coordinate $\xi_1$:

$$\alpha = \begin{cases} \alpha_\mu \\ \alpha_m \\ \alpha_l \end{cases}, \quad A_{\mu l} = \begin{cases} A_\mu(1 + \Delta_\mu \xi_1^{-\varepsilon_\mu}) \\ A_m \psi^{\delta_m} \\ A_l \psi^{\delta_l}(1 + \Delta_l \xi_1^{\varepsilon_l}) \end{cases} \quad \text{for } \xi_1 = \begin{cases} \xi_1 \geq 2 \\ 0.01 \leq \xi_1 \leq 2. \\ \xi_1 \leq 0.01 \end{cases} \tag{A17}$$

Herein,

$$\xi_1 = \frac{\xi}{\psi^{1/\varepsilon_\mu}}, \, \varepsilon_\mu = \alpha_\mu - (1 - \beta_l), \, \varepsilon_l = \alpha_l - (1 - \beta_l), \, \delta_l = \frac{\varepsilon_\mu - \varepsilon_l}{\varepsilon_\mu}, \, \delta_m = \frac{\alpha_\mu - \alpha_m}{\varepsilon_\mu}, \tag{A18}$$

$$\Delta_\mu = \frac{1}{A_\mu}\frac{n}{n+1+C_{1-\beta_l/C_\mu}}, \, \Delta_l = A_l\frac{n}{2n+1+C_{\alpha_l+\varepsilon_l/C_{\alpha_l}}}, \, C_\gamma = (1 - \gamma)B(\gamma),$$

with $B(\gamma)$ given by (3.12), while the parameters $\alpha_m$ and $A_m$ are defined by linear interpolation of the values of $\Omega_1(\xi_1)$ in log-log coordinates between the points $\xi_1 = 0.01$ and $\xi_1 = 2$:

$$\alpha_m = \frac{\log\Omega_1(2) - \log\Omega_1(0.01)}{2.3010}, \, A_m = \frac{\Omega_1(2)}{2^{\alpha_m}} \tag{A19}$$

with $\Omega_1(2) = 2^{\alpha_\mu}A_\mu(1 + \Delta_\mu 2^{-\varepsilon_\mu})$ and $\Omega_1(0.01) = 0.01^{\alpha_l}A_l(1 + \Delta_l 0.01^{\varepsilon_l})$.

For certainty, it is assumed that $\alpha_l < \alpha_\mu$. The error of the approximate solution (A16), A17) does not exceed 5% in the entire range of the parameter $\psi$ ($0 \leq \psi < \infty$) and coordinate $\xi$ ($0 \leq \xi < \infty$).

In the particular case of a Newtonian fluid ($n = 1$) and Carter's leak-off ($\beta_l = 1/2$), the definitions of $\alpha_\mu$, $\alpha_l$, $A_\mu$, $A_l$ and equations (A18), (A19) yield: $\alpha_\mu = 2/3$, $\alpha_m = 0.6484$, $\alpha_l = 5/8$, $A_\mu = 3.147$, $A_m = 3.634$, $A_l = 2.534$, $\varepsilon_\mu = 1/6$, $\varepsilon_l = 1/8$, $\delta_m = 0.10962$, $\delta_l = 1/4$, $\Delta_\mu = 0.1589$, $\Delta_l = 0.5138$.

# References

Abou-Sayed A.S., Thompson T.W. & Keckler K. 1994 Safe injection pressures for disposing of liquid wastes: a case study for deep well injection (SPE/ISRM-28236). *Proceedings of the Second SPE/ISRM Rock Mechanics in Petroleum Engineering*, Balkema, 769–776.

Adachi J.I. & Detournay E. 2002 Self-similar solution of plane-strain fracture driven by a power-law fluid. *Int. J. Numer. Anal. Meth. Geomech*., **26**, 579-604.

Adachi J., Siebrits E., Pierce A. & Desroches J. 2007 Computer simulation of hydraulic fractures. *Int. J. Rock Mech. Mining Sci.,* **44**, 739-757.

Bolzon G. & Cochetti G. 2003 Direct assessment of structural resistance against pressurized fracture. *Int. J. Numer. Anal. Methods Geomech.,* **27**, 353–378.

Bunger A.P. 2005. Near-surface hydraulic fracture. *Ph.D. Thesis*, University of Minnesota.

Crowe C.T., Elger D.F., Williams B.C. & Roberson J.A. 2009 *Engineering Fluid Mechanics*. 9[th] ed., John Willey & Sons, Inc.

Descroches J., Detournay E., Lenoach B., Papanastasiou P., Pearson J.R.A., Thiercelin M. & Cheng A. 1994 The crack tip region in hydraulic fracturing. *Proc. Roy Soc. London, Ser. A*, **447**, 39-48.

Economides M. & Nolte K. 2000 *Reservoir Simulation*. 3[rd] edn. Wiley, Chichester, UK.

*Effective and Sustainable Hydraulic Fracturing* 2013 A.P. Bunger, J. McLennan, R. Jeffrey (eds), Published by InTech (Croatia). Proc. Int. Conference HF-2013. Available on line: www.intechopen.com.

Garagash D.I. 2006 Propagation of a plane-strain hydraulic fracture with a fluid lag: Early time solution. *Int. J. Solids Struct*. **43**, 5811-5835.

Garagash D.I. & Detournay E. 2000 The tip region of a fluid-driven fracture in an elastic medium. *ASME J. Appl. Mech.*, **67**, 183-192.

Garagash D.I., Detournay E. & Adachi J.I. 2011 Multiscale tip asymptotics in hydraulic fracture with leak-off. *J. Fluid Mech.*, **669**, 260-297.

Geertsma J. & de Klerk F. 1969 A rapid method of predicting width and extent of hydraulically induced fractures. *J. Pet. Tech*., **21**, 1571-1581.

Gordeliy E. & Detournay E. 2011 A fixed grid algorithm for simulating the propagation of a shallow hydraulic fracture with a fluid lag. *Int. J. Numer. Anal. Methods Geomech*. **35**(5), 602–629.

Gordeliy E. & Peirce A. 2013 Coupling schemes for modeling hydraulic fracture propagation using the XFEM. *Computer Meth. Appl. Mech. Eng*., **253**, 305–322.

Jeffrey R.G. & Mills K.W. 2000 Hydraulic fracturing applied to inducing longwall coal mine goaf falls. *Pacific Rocks 2000*. Rotterdam, Balkema, 423–430.

Kemp L.F. 1990 Study of Nordgren's equation of hydraulic fracturing. *SPE Production Eng*., 5 (SPE 19959), 311-314.

Khristianovich S.A. & Zheltov V.P. 1955 Formation of vertical fractures by means of highly viscous liquid. In: *Proc. 4-th World Petroleum Congress*, Rome, 579-586.

Lecampion B. & Detournay E. 2007 An implicit algorithm for the propagation of a hydraulic fracture with a fluid lag. *Comput. Methods Appl. Mech. and Engng,* **196**(49–52), 4863–4880.

Lenoach B. 1995 The crack tip solution for hydraulic fracturing in a permeable solid. *J. Mech. Phys. Solids*, **43**, 1025-1043.

Linkov A.M. 2011 Speed equation and its application for solving ill-posed problems of hydraulic fracturing. *Doklady Physics*, **56**, 436-438 [Translation from Russian: Линьков А.М. 2011 Уравнение скорости и его применение для решения некорректных задач о гидроразрыве. *Доклады Академии наук*, **439**, 473-475].

Linkov A.M. 2012 On efficient simulation of hydraulic fracturing in terms of particle velocity. *Int. J. Eng. Sci.,* **52**, 77-88.

Linkov A.M. 2014 Universal asymptotic umbrella for hydraulic fracture modeling. Available online at http://arxiv.org/abs/1404.4165 Date: Wed, 16 Apr 2014 08:35:16 GMT   (564kb) Cite as: arXiv: 1404.4165 [physics.flu-dyn]. 11 p.

Linkov A.M. & Mishuris G. 2013 Modified formulation, ε-regularization and efficient solving of hydraulic fracture problem. *Effective and Sustainable Hydraulic Fracturing*. A.P. Bunger, J. McLennan, R. Jeffrey (eds), Published by InTech (Croatia), p. 641-657. Proc. Int. Conference HF-2013. Available on line: www.intechopen.com.

Lister J.R. 1990 Buoyancy-driven fluid fracture: the effects of material toughness and of low-viscosity precursors. *J. Fluid Mech*., **210**, 263–280.

Mishuris G., Wrobel M. & Linkov A.M. 2012 On modeling hydraulic fracture in proper variables: stiffness, accuracy, sensitivity. *Int. J. Eng. Sci*., **61**, 10-23.

Mitchell S.L., Kuske R. & Pierce A.P. 2007 An asymptotic framework for analysis of hydraulic fracture: the impermeable fracture case. *ASME J. Appl. Mech*., **74**, 365-372.

Murdoch L.C. & Slack W.W. 2002 Forms of hydraulic fractures in shallow fine-grained formations. *J. Geotech. Geoenvironmental Engng,* **128**(6), 479–487.

Muskhelishvili N.I. 1953 *Singular Integral Equations*. Noordhoff, Groningen.

Muskhelishvili N.I. 1975 *Some Basic Problems of the Mathematical Theory of Elasticity*. Noordhoff, Groningen.

Nordgren R.P. 1972 Propagation of a vertical hydraulic fracture. *Soc. Pet. Eng. J*. August, 306-314.

Peirce A. & Detournay E. 2008 An implicit level set method for modeling hydraulically driven fractures. *Comput. Methods Appl. Mech. Engng., **197***, 2858-2885.

Penman A.D.M. 1977 The failure of the Teton dam. *Ground Engineering, **10***(6), 18–27.

Perkins K. & Kern L.F. 1961 Widths of hydraulic fractures. *J. Pet. Tech*., **13**, 937-949.

Pollard D.D. & Johnson A.M. 1973 Mechanics of growth of some laccolithic intrusions in the Henry Mountains, Utah. II. *Tectonophysics, **18***, 311–354.

Rice J.R. 1968 Mathematical analysis in the mechanics of fracture, in: H. Liebowitz (ed.), *Fracture, an Advanced Treatise*, vol. II, Academic Press, New York, NY, 191–311 (Chapter 3).

Roper S.M. & Lister J.R. 2005 Buoyancy-driven crack propagation from an over-pressured source. *Journal of Fluid Mechanics, **536***, 79–98.

Rubin A.M. 1995 Propagation of magma-filled cracks. *Annual Review of Earth and Planetary Sciences, **23***, 287–336.

Savitski A. & Detournay E. 2002 Propagation of a fluid driven penny-shaped fracture in an impermeable rock: asymptotic solutions. *Int. J. Solids Struct*., **39**, 6311-6337.

Sethian J.A. 1999 *Level Set Methods and Fast Marching Methods*. Cambridge, Cambridge University Press.

Spence D.A. & Sharp P.W. 1985 Self-similar solutions for elastohydrodynamic cavity flow. *Proc. Roy Soc. London, Ser. A*, **400**, 289-313.

Spence D.A. & Turcotte D.L. 1985 Magma-driven propagation crack. *J. Geophys. Res*., **90**, 575–580.

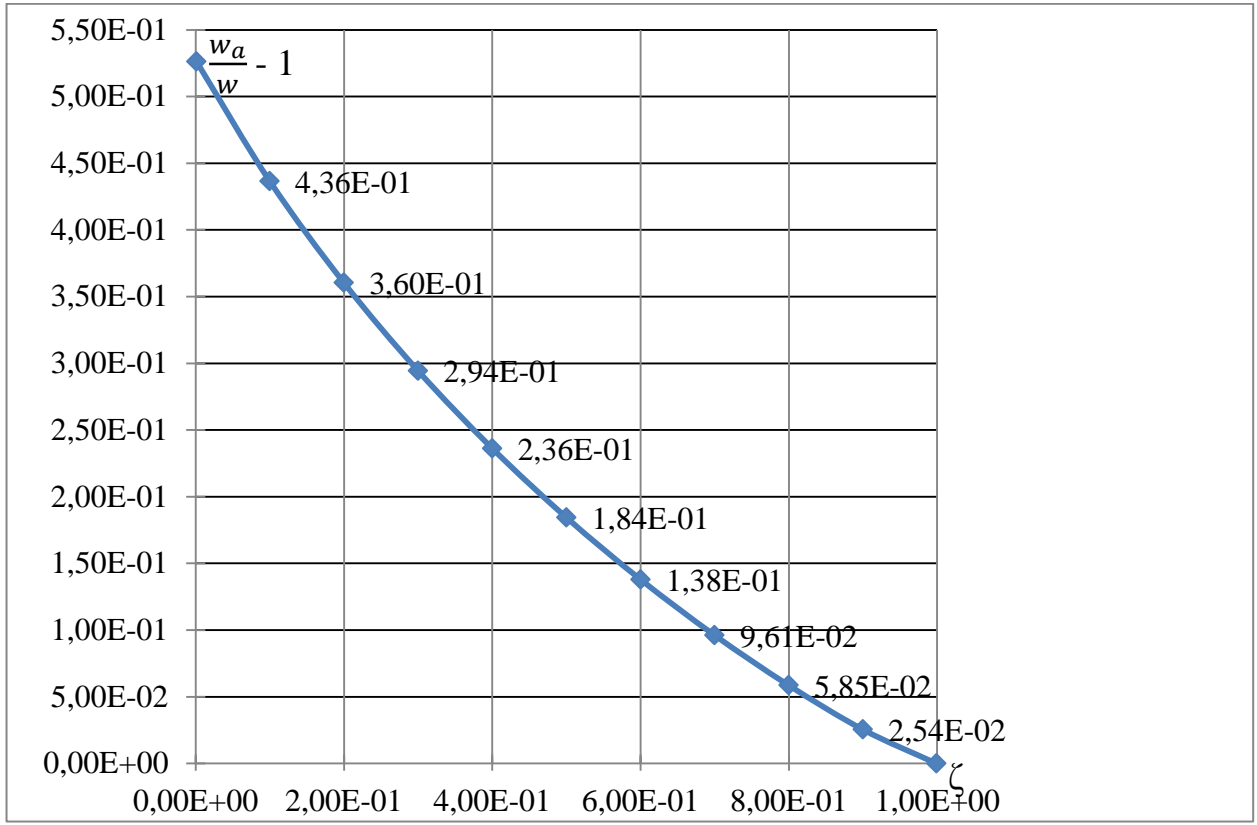

Fig. 1. Relative error of asymptotic opening

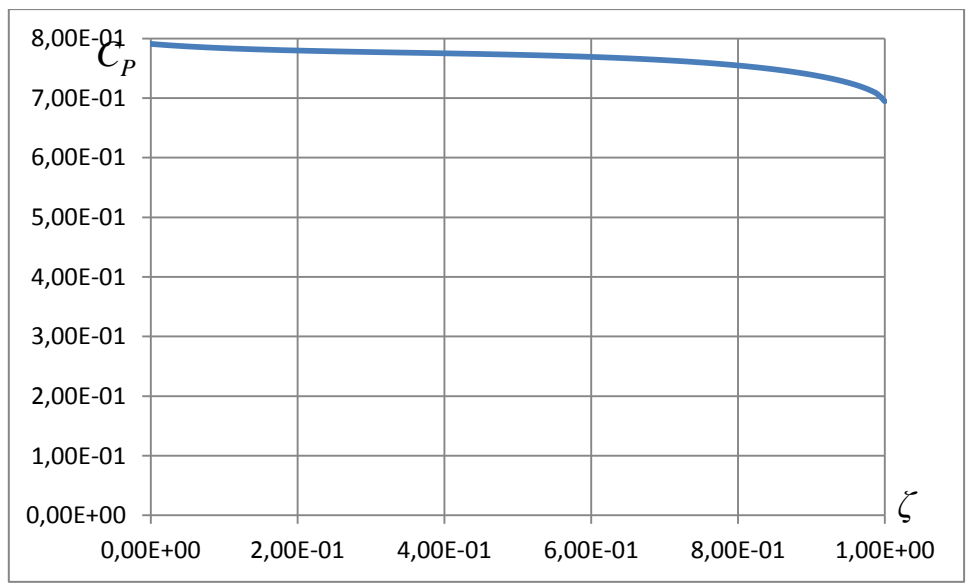

Fig. 2. Self-similar matching pressure constant